\documentclass[twocolumn,runningheads]{svjour2}
\smartqed  
\usepackage{graphicx}
\usepackage{mathptmx}   
\usepackage[authoryear]{natbib}
\journalname{Astrophysics and Space Science (CoRoT/ESTA Volume)}
\begin{document}

\title{The Li\`ege Oscillation Code}
\author{R.~Scuflaire \and J.~Montalb\'an \and S.~Th\'eado \and P.-O.~Bourge
  \and A.~Miglio \and M.~Godart \and A.~Thoul \and A.~Noels}
\institute{R.~Scuflaire, S.~Th\'eado, J.~Montalban \and P.-O.~Bourge,
  \and A.~Miglio \and M.~Godart \and A.~Thoul \and A.~Noels
  \at
  Institut d'Astrophysique et de G\'eophysique, Universit\'e de Li\`ege, all\'ee
  du 6 Ao\^ut 17, B-4000 Li\`ege, Belgium
}

\date{Received: date / Accepted: date}

\maketitle
\begin{abstract}
The Li\`ege Oscillation code can be used as a stand-alone program or as a
library of subroutines that the user calls from a Fortran main program of his
own to compute radial and non-radial adiabatic oscillations of stellar models.
We describe the variables and the equations used by the program and the methods
used to solve them. A brief account is given of the use and the output of the
program.  
\end{abstract}

\section{Introduction}\label{intro}

The Li\`ege oscillation code (OSC) has been developed in the early 70s for
computing adiabatic pulsations of sperically symmetric stars (no rotation nor
magnetic field).  It has gone through minor updates and is still presently in
use in the asteroseismology group of the Li\`ege Institute of Astrophysics and
Geophysics. Besides the frequencies and eigenfunctions, it produces also the
coefficients needed to compute the first order rotational frequency splitting
for a rigid rotation and the kernels needed to compute the splitting when a non
rigid rotation is considered.

\section{Stellar models}\label{sec:models}

A stellar model is input to OSC as a table describing a few physical
quantities at discrete points of the star, ordered from the centre to the
surface. In theory, only two functions are necessary to compute stellar
oscillations, for instance $\rho(r)$ and $\Gamma_1(r)$, the density and the
first adiabatic exponent in terms of the radius. However, for OSC, the model
file must give at each point the values of the radius $r$, the mass $m$ in the
spere of radius $r$, the total pressure $P$, the density $\rho$ and the first
adiabatic exponent $\Gamma_1$. It is clear that $m$ and $P$ could have been
computed from the other quantities. The program does not require the
Brunt-V\"ais\"al\"a frequency, often poorly computed by evolution codes.

If the first point is not at the centre, OSC computes the oscillations of the
given envelope with a rigid boundary condition at the bottom. Of course, when
neglecting the oscillatory behaviour of the core, great attention must be paid
to the physical meaning of the output of the program.

The outer boundary conditions will be applied at the last point, considered as
the surface of the star.

Inside the code, the model is described by the following five dimensionless
quantities: $x=r/R$, $q/x^3$ (with $q=m/M$), $RP/GM\rho$, $4\pi R^3\rho/M$ and
$\Gamma_1$, where $R$ and $M$ denote the radius and the total mass of the star.


\section{Stellar oscillations}\label{sec:oscill}

\subsection{Oscillation modes}

The small perturbations of a spherical star without rotation or magnetic
field may be described as a superposition of normal modes of oscillation which
are the solutions of a linear boundary eigenvalue problem. These normal modes
may be indexed by three integers $k$, $\ell$ and $m$.  Index $k$ is loosely
related to the number of nodes of the radial displacement.  Indices $\ell$ and
$m$ are the usual indices of the spherical function $Y_{\ell m}(\theta,\phi)$
describing the angular dependence.  Index $\ell$ may take any null or positive
integer value and $m$ may take $2\ell+1$ values between $-\ell$ and $+\ell$.

In the following description we use the notation $\delta X$ and $X'$ for
the Lagrangian and Eulerian perturbation of any quantity $X$ and
$\sigma$ for the angular frequency. We often use the dimensionless
angular frequency $\omega=\sigma\tau_{dyn}$, where the dynamical time
$\tau_{dyn}$ is defined as $\sqrt{R^3/GM}$.

\subsection{Oscillation equations}

The theory of stellar oscillation has been developed in a number of textbooks.
We are the most familiar with the paper of \citet{L1958} and the books of
\citet{U1979} and \citet{C1980}. We will just write the needed equations in the
form they are implemented in our code.

\subsubsection{Radial oscillations.}

In the case of radial oscillations ($\ell=m=0$), the equation of Poisson can be
integrated and the perturbation of the gravitational potential eliminated.  The
differential system is then reduced to order two. We describe a normal mode
with two functions $Y(x)$ and $Z(x)$.  Disregarding an arbitrary phase, the
displacement \vec{\delta r} and the lagrangian perturbation of the pressure
$\delta P$ are written in terms of $Y(x)$ and $Z(x)$ in the following way.
\begin{equation}
\vec{\delta r}=\Re\{a(r)e^{-i\sigma t}\vec{e}_r\}
=\sqrt{4\pi}\Re\{a(r)Y_{00}(\theta,\phi)e^{-i\sigma t}\vec{e}_r\}\,,
\end{equation}
where $\vec{e}_r$ is a unit vector in the radial direction.
Near the centre, $a(r)\propto r$ and may be written
\begin{equation}
a(r)/r=Y(x)\quad\mbox{or}\quad a(r)/R=xY(x)\,.
\end{equation}
In a similar way,
\begin{equation}
\delta P/P=\Re\{Z(x)e^{-i\sigma t}\}
=\sqrt{4\pi}\Re\{Z(x)Y_{00}(\theta,\phi)e^{-i\sigma t}\}\,.
\end{equation}
Now, the differential equations read
\begin{eqnarray}
\frac{dY}{dx}&=&-\frac{3}{x}Y-\frac{1}{\Gamma_1x}Z\,,\\
\frac{dZ}{dx}&=&\left(4\frac{q}{x^3}+\omega^2\right)\frac{GM\rho}{RP}xY
  +\frac{GM\rho}{RP}\frac{q}{x^3}xZ\,.
\label{eq:radial_dZ/dx}
\end{eqnarray}
These equations must be completed by the boundary conditions. At the centre, the
regularity of the solution is ensured by the condition
\begin{equation}
3\Gamma_1Y+Z=0\quad\mbox{at}\quad x=0\,.
\end{equation}

When an envelope model is given, the condition at the centre is replaced by
a condition at the bottom of the envelope.
\begin{equation}
Y=0\,.
\end{equation}

At the surface, we generally apply a condition deduced from the vanishing of the
pressure. In this case, the coefficient $GM\rho/RP$ in equation
(\ref{eq:radial_dZ/dx}) tends to infinity and the regularity of the solution
requires that
\begin{equation}
\left(4\frac{q}{x^3}+\omega^2\right)Y+\frac{q}{x^3}Z=0\,.
\end{equation}
Note that if $R$ is the value of $r$ at the last point, $x=q/x^3=1$ at the
surface but this is not mandatory.  The user can choose to apply the more usual
condition
\begin{equation}
\delta P= 0\quad\mbox{or}\quad Z=0\,.
\end{equation}

\subsubsection{Non radial oscillations.}

In the case of non radial oscillations ($\ell\ne0$), the differential
system is of order four. We describe a normal mode with four functions
$Y(x)$, $Z(x)$, $U(x)$ and $V(x)$. These functions as well as the
frequency do not depend upon index $m$ ($2\ell+1$-fold degeneracy). 
The displacement reads
\begin{eqnarray}
\overrightarrow{\delta r}&=&\sqrt{4\pi}\Re\left\{\left[a(r)
Y_{\ell m}(\theta,\phi)\vec{e}_r
+b(r)\left(\frac{\partial Y_{\ell m}(\theta,\phi)}{\partial\theta}\vec{e}_\theta
\right.\right.\right.\nonumber \\
& &\left.\left.\left.+{1\over\sin\theta}
\frac{\partial Y_{\ell m}(\theta,\phi)}{\partial\phi}
\vec{e}_\phi\right)\right]e^{-i\sigma t}\right\}\,,
\end{eqnarray}
where $\vec{e}_r$, $\vec{e}_\theta$ and $\vec{e}_\phi$ form the usual
local cartesian basis of spherical coordinates.
Near the centre, $a(r)$ and $b(r)\propto r^{\ell-1}$ and are written
\begin{eqnarray}
a(r)/R&=&x^{\ell-1}Y(x)\,,\\
b(r)/R&=&{x^{\ell-1}\over\omega^2}\left[U(x)+{RP\over GM\rho}Z(x)
+{q\over x^3}Y(x)\right]\,.
\end{eqnarray}
The Lagrangian perturbation of pressure $\delta P$ and the Eulerian
perturbation of the gravitational potential $\Phi'$ are given by
\begin{eqnarray}
&&{\delta P\over P}=\sqrt{4\pi}\Re\{x^\ell Z(x) Y_{\ell m}(\theta,\phi)
e^{-i\sigma t}\}\,,\\
&&{R\Phi'\over GM}=\sqrt{4\pi}\Re\{x^\ell U(x) Y_{\ell m}(\theta,\phi)
e^{-i\sigma t}\}\,,\\
&&{R^2\over GM}{\partial\Phi'\over\partial r}=\sqrt{4\pi}\Re\left\{
x^{\ell-1}\left[ V(x)-{4\pi R^3\rho\over M}Y(x)\right]\right.\nonumber \\
&&\qquad \times Y_{\ell m}(\theta,\phi)e^{-i\sigma t}\biggr\}\,.
\label{defV}
\end{eqnarray}
Of course, the solution of a linear problem may be multiplied by an
arbitrary factor and the $\sqrt{4\pi}$ factor in the above expressions
may be dropped. We have put it there for aesthetic reasons because the
spherical functions are normalized in such a way that
\begin{equation}
\int_{4\pi}|Y_{\ell m}|^2d\Omega=1\,.
\end{equation}
With this $\sqrt{4\pi}$ factor, the time-average kinetic energy of a
mode is given by
\begin{eqnarray}
\overline{E}_{kin}&=&\int{1\over2}\rho\overline{v^2}\,dV
={\sigma^2\over4}\int[a^2+\ell(\ell+1)b^2]4\pi r^2\rho\,dr \nonumber \\
&=&{\sigma^2\over4}\int[a^2+\ell(\ell+1)b^2]\,dm\,,
\end{eqnarray}
without any $\pi$ factor.

With these definitions, the oscillation equations read
\begin{eqnarray}
\frac{dY}{dx}&=&\frac{\ell+1}{x}\left\{-Y+\frac{\ell}{\omega^2}\left(
  \frac{q}{x^3}Y+\frac{RP}{GM\rho}Z+U\right)\right\}\nonumber\\
  &&-\frac{x}{\Gamma_1}Z\,,\\
\frac{dZ}{dx}&=&\frac{GM\rho}{RP}\biggl\{\left(\omega^2+4\frac{q}{x^3}\right)
  \frac{Y}{x}+x\frac{q}{x^3}Z-\frac{V}{x}\nonumber\\
  &&-\frac{\ell(\ell+1)}{x\omega^2}
  \frac{q}{x^3}\left(\frac{q}{x^3}Y+\frac{RP}{GM\rho}Z+U\right)\biggr\}
  -\frac{\ell}{x}Z\,,\\
\frac{dU}{dx}&=&\frac{1}{x}\left(V-\frac{4\pi R^3\rho}{M}Y-\ell U\right)\,,\\
\frac{dV}{dx}&=&\frac{\ell+1}{x}(\ell U-V)\nonumber\\
  &&+\frac{\ell(\ell+1)}{x\omega^2}
  \frac{4\pi R^3\rho}{M}\left(\frac{q}{x^3}Y+\frac{RP}{GM\rho}Z+U\right)\,.
\end{eqnarray}
These equations have been published by \citet{B1975}, but their equation~(9) has
been affected by a typo (an extra factor $\ell$).

The regularity of the solution at the centre imposes two conditions at $x=0$.
\begin{eqnarray}
Y&=&\frac{\ell}{\omega^2}\left[\frac{q}{x^3}Y+\frac{RP}{GM\rho}Z+U\right]\,,\\
V&=&\frac{4\pi R^3\rho}{M}Y+\ell U\,.
\end{eqnarray}

In the case of an envelope, the bottom of the envelope is supposed to behave as
a rigid boundary.
\begin{equation}
Y=0\,.
\end{equation}
There are no movement nor density perturbations in the core, where the
perturbation of the gravitational potential obeys a Laplace equation and has a
simple analytical expression. We thus require a continuous match between $\Phi'$
and its gradient at the bottom of the envelope.  This is expressed as
\begin{equation}
V=\ell U\,.
\end{equation}

Two boundary conditions must be imposed at the surface. The first one involves
the lagrangian perturbation of the pressure $\delta P$.  As for the radial case,
the default choice is deduced from the requirement of regularity of the solution
when $P$ vanishes at the surface. In our variables, this condition reads
\begin{eqnarray}
\lefteqn{\left[\omega^2+4\frac{q}{x^3}-\frac{\ell(\ell+1)}{\omega^2}
\left(\frac{q}{x^3}\right)^2\right]\frac{Y}{x}+x\frac{q}{x^3}Z}\nonumber \\
&&\quad-\frac{\ell(\ell+1)}{\omega^2x}\frac{q}{x^3}U-\frac{V}{x}=0\,.
\end{eqnarray}
The user can however choose to impose the more usual condition
\begin{equation}
\delta P=0\quad\mbox{or}\quad Z=0\,.
\end{equation}

The second boundary condition ensures the matching of $\Phi'$ and its
gradient with the regular solution of the Laplace equation outside the star,
\begin{equation}
V+(\ell+1)U=0\,.
\label{2ndBC}
\end{equation}

Our choice of variables and the way the differential equations are written call
for three remarks.\\
(1) The inclusion of a term in $Y$ in the definition of $V$
(equation~\ref{defV}) ensures that the boundary condition~(\ref{2ndBC}) is still
valid in the (unphysical) case of a non vanishing density at the surface of the
model.\\
(2) The use of the lagrangian perturbation of the pression results in a better
precision in the external layers.\\
(3) We do not use the Brunt-V\"ais\"al\"a frequency $n$ in the coefficients of
the equations, often badly computed by stellar evolution codes.\\
However, we use non independent functions $\rho(r)$, $m(r)$ and $P(r)$. Troubles
may stem from their possible inconsistencies. Maybe it would have been wiser to
let OSC compute $m$ and $P$ from $\rho$.

The solutions computed by OSC are normalized in such a way that
$$\int[a^2+\ell(\ell+1)b^2]\,dm=MR^2\,.$$

\subsection{Mode classification}
The radial modes owe their existence to the compressibility of the stellar
material (acoustic or pressure modes). The mode with the lowest frequency is
called the fundamental mode, it has no node in the displacement (except at the
origin) and is called $p_1$. By order of increasing frequency and number of
nodes, we have then the first harmonics (one node, $p_2$), the second harmonics
(two nodes, $p_3$), \dots

For non radial modes, the situation is more complicated. For each
$\ell$, we have a spectrum of $p$-modes. They are of the same nature as
the radial modes, owing their existence to the compressibility of the
stellar material. They are numbered $p_1$, $p_2$, \dots by order of
increasing frequency. If the stellar model has a radiative zone, it has
a spectrum of $g^+$-modes. Their frequencies are lower than those of the
$p$-modes and have an accumulation point at zero. They are numbered
$g^+_1$, $g^+_2$, \dots by order of decreasing frequency. They owe their
existence to the buoyancy force. For each value of $\ell>1$, there exists
one $f$-mode, with its frequency between those of the $p$-modes and the
$g$-modes. This mode does not disappear when the stellar material is
incompressible nor when the buoyancy force is zero. When the star has a
convective zone, another spectrum appears, the $g^-$-modes. They have an
exponential temporal behaviour (their frequencies are imaginary) and are
associated with convection. We neglect them in the following discussion.

In OSC, we use an integer to denote the type and order of a computed mode: $n$
for $p_n$, $-n$ for $g^+_n$ and $0$ for $f$. In the Cowling's approximation (the
Eulerian perturbation of the gravitational potential is neglected), the order of
a given mode can be easily deduced from the behaviour of the vertical and
horizontal displacements \citep{S1974b, G1979, G1980}. In our case, the mode
number obtained in the same way is generally correct for models which are not
too evolved. But as the condensation (measured by the ratio $\rho_c/\bar{\rho}$)
increases with the age of the model, it can just be considered as a clue and
finally looses any meaning. The algorithm described by \citet{L1985} gives a
clue to the mode number (also computed by OSC) which keeps its utility a bit
longer. But when the condensation of the model is really too high, the only
reliable identification method consists in the computation of a large number of
contiguous modes, up to the asymptotic domain, where the implemented algorithms
continue to give reliable mode numbers.

Though the order of the mode cannot be obtained safely, its parity can and is
provided by OSC.

\subsection{Influence of rotation}
The rotation of the star removes the degeneracy of the non radial oscillation
frequencies. If $\sigma_{k\ell}^0$ denotes the frequency in the absence of
rotation, a slow solid rotation with angular velocity $\Omega$ slightly alters
the frequencies in the following way
\begin{equation}
\sigma_{k\ell m}=\sigma_{k\ell}^0+m\beta_{k\ell}\Omega\,,
\end{equation}
with
\begin{equation}
\beta_{k\ell}=1-{\displaystyle{\int(b^2+2ab)\,dm}
\over\displaystyle{\int[a^2+\ell(\ell+1)b^2]\,dm}}\,.
\end{equation}

When the angular velocity depends on the radius, the altered frequencies may be
written
\begin{equation}
\sigma_{k\ell m}=\sigma_{k\ell}^0+m\int K_{k\ell}(x)\Omega(x)\,dx\,,
\end{equation}
where the kernel $K_{k\ell}(x)$, computed by OSC, is given by
\begin{equation}
K_{k\ell}(r)=
\frac{\rho r^2\left[a^2+\ell(\ell+1)b^2-2ab-b^2\right]}
{\displaystyle\int\rho r^2\left[a^2+\ell(\ell+1)b^2\right]\,dr}\,.
\end{equation}

\subsection{Physical description of the modes}
For low order modes, specially for evolved models, the physical characteristics
of a mode is not tightly linked to its $g$ or $p$ label \citep{S1974a,S1980}.
OSC outputs different indexes allowing the user a quick analysis
of the physical behaviour of a computed mode (gravity or pressure wave,
trap\-ped mode, \dots).

\section{Technique of solution}\label{sec:solution}

\subsection{Interpolation of the model}
The grid of points used for the computation of the model is rarely appropriate
for the computation of oscillations, as the eigenfunctions can exhibit rapid
spatial oscillations in regions where the variables describing the model are
well-behaved. As the oscillatory behaviour of eigenfunctions is easy to
foresee, we interpolate the model before any oscillation computation, increasing
the number of points where they will prove necessary. The interpolation method
we use preserves the continuity of the first derivatives.

\subsection{Difference equations}
We have adopted a difference equation scheme of the fourth order. That is why we
do not need to use Richardson extrapolation method to increase the precision of
the eigenfrequency.

Our difference scheme rests on the following identity satisfied by any vector
function $\vec{y}(x)$ with continuous derivatives up to the fifth order.
\begin{equation}
\vec{y}_i+\frac{h}{2}\vec{y}'_i+\frac{h^2}{12}\vec{y}''_i=
\vec{y}_{i+1}-\frac{h}{2}\vec{y}'_{i+1}+\frac{h^2}{12}\vec{y}''_{i+1}
+O(h^5)\,,
\label{eq:identity}
\end{equation}
where $\vec{y}_i$ and $\vec{y}_{i+1}$ are the values of $\vec{y}(x)$ at points
$x_i$ and $x_{i+1}$ and $h=x_{i+1}-x_i$.
If $\vec{y}$ is a solution of the linear differential system
\begin{equation}
\frac{d\vec{y}}{dx}=A(x)\vec{y}\,,
\end{equation}
identity (\ref{eq:identity}) may be written
\begin{eqnarray}
\left\{1+\frac{h}{2}\alpha_i+\frac{h^2}{12}\beta_i\right\}\vec{y}_i=
\left\{1-\frac{h}{2}\alpha_{i+1}+\frac{h^2}{12}\beta_{i+1}\right\}\vec{y}_{i+1}
  \nonumber\\
\quad+O(h^5)\,,
\end{eqnarray}
with
\begin{eqnarray}
\alpha&=&A\,, \\
\beta&=&A^2+\frac{dA}{dx}\,.
\end{eqnarray}
The difference equations are easily obtained from the above equations,
neglecting the term $O(h^5)$. At the centre, certain coefficients of the matrix
$A$ are singular and a slightly different treatment is needed.  The matrix can
be written as
\begin{equation}
A(x)=\frac{1}{x}B(x)\,,
\end{equation}
with all the odd order derivatives of matrix $B$
vanishing at $x=0$. It is clear that the regularity of the solution requires, at
$x=0$,
\begin{equation}
B\vec{y}=0\,.
\label{eq:boundary}
\end{equation}
It is worth noticing that the rank of $B(0)$ is lower than its dimension and
that equation~(\ref{eq:boundary}) gives the right number of boundary conditions
(1 for the radial case and 2 for the nonradial one). The matrices $\alpha$ and
$\beta$ assume the following different forms at the centre (index $0$).
\begin{eqnarray}
\alpha&=&0\,, \\
\beta&=&(2-B_0)^{-1}\left(\frac{d^2B}{dx^2}\right)_0\,.
\end{eqnarray}

\subsection{Inverse iteration method}

After the discretization of the differential equations we are left with an
algebraic eigenvalue problem where the eigenvalue is $\lambda=\omega^2$. OSC
uses the inverse iteration method.  It is a powerful tool in linear eigenvalue
problems. It is described in the book of \citet[chap~9, sect~47]{W1965}. It has
been used by \citet{K1977} to compute stellar radial nonadiabatic oscillations.
The principle of the method is easy to describe.  Consider the following
eigenvalue problem
\begin{equation}
(A-\lambda B)\vec{y}=0\,.
\label{eq:eigen}
\end{equation}
The method is generally exposed with a unit matrix in place of B. Suppose that
we know an approximation $\lambda_0$ of an eigenvalue $\lambda$. Starting with
an arbitrary vector $\vec{y}_0$, we build the sequence
\begin{equation}
\vec{y}_{n+1}=-(A-\lambda_0B)^{-1}B\vec{y}_n\quad n=0,1,2,\ldots
\end{equation}
Then,
\begin{equation}
\frac{\vec{y}_n\cdot\vec{y}_{n+1}}{\vec{y}_{n+1}\cdot\vec{y}_{n+1}}
\rightarrow\lambda-\lambda_0\,.
\end{equation}
The convergence is quite fast. Practically, the $\vec{y}_n$ must be normalized
at each step of the computation to avoid overflow.  Moreover, these normalized
$\vec{y}_n$ tend to the eigenvector associated with $\lambda$.

It is true that, in the case of nonradial oscillations, the problem to solve is
not exactly in the form of equation~(\ref{eq:eigen}). But it is put in the right
form if we write
\begin{equation}
\lambda=\lambda_0+\Delta\lambda
\end{equation}
and linearize with respect to the correction $\Delta\lambda$. The solution is
then obtained in a few iterations of this process.

\section{Use of the program}\label{sec:usage}

OSC is written in Fortran and can be used either as a stand-alone program or as
a library of subroutines that the user calls from a main program of his own.
The stand-alone program is in fact just a user interface to the library. It
accepts instructions from the standard input (or from a command file) and prints
all kind of information to the standard output. At the request of the user, the
eigenfunctions can be saved to a file. For heavy work, the user had better write
his own Fortran main program and call the routines of the library. He gains a
better control on the computation and acces to results not available otherwise
(the rotation kernels for a $r$-dependent angular velocity $\Omega$, for
instance).

\section{Applications}

OSC is routinely used in our group in Li\`ege and by members of the Belgian
Asterosesismology Group (BAG) for seismic studies of solar-like pulsators such
as, e.g., $\alpha$~Cen~A+B \citep{T2003a,M2005} and of classical $\beta$~Cephei
variables \citep{A2003,T2003b,D2004,A2004,T2006,B2007}.  It is worth mentioning
that in these studies, we obtained indications on the internal rotation of the
$\beta$~Cephei stars HD~129929 and $\theta$~Ophiuchi.

The adiabatic frequencies computed by OSC are also used as first approximations
by the program MAD which computes the non adiabatic oscillations of stellar
models. See \citet{D2001} for a description of the code.

The Li\`ege oscillation code has also taken part in the work and code
comparisons realized within the Corot/ESTA group \citep{M2007a, M2007b}.

\section{Discussion}

There is not any general agreement on what the outer boundary condition on the
perturbation of pressure should be and different boundary conditions are
implemented in the existing codes.  In our opinion, the precise choice of the
outer boundary condition does not matter so much, as, in any way, the
oscillation is far from adiabatic in the very external layers of the star.
Moreover, \citet{D200} have shown that in the Sun, waves in the frequency range
$\nu\approx2-10$~mHz may reach the chromosphere-corona transition regions by
means of a tunneling through the atmospheric barrier. When comparing with
observations, a reasonable strategy is thus to use expressions of frequencies
which are insensitive to the very external stellar layers \citep{R2003,R2005}.

\begin{acknowledgements}
We acknowledge financial support from the Belgian Science Policy Office (BELSPO)
in the frame of the ESA PRO\-DEX~8 program (contract C90199), from the Belgian
Interuniversity Attraction Pole (grant P5/36) and from the Fonds National de la
Recher\-che Scientifique (FNRS).
\end{acknowledgements}


\end{document}